\newtheorem{postulate}{Postulate}
\newtheorem{theorem}{Theorem}
\documentclass[12pt]{article}

\font\gothic=eufm10 at 12pt
\font\blackboard=msbm10 at 12pt
\def\bbf#1{\hbox{\blackboard #1}}
\def\goth#1{\hbox{\gothic #1}}

\def\cA{{\cal A}}

\def\cD{{\cal D}}
\def\cG{{\cal G}}
\def\cH{{\cal H}}
\def\cL{{\cal L}}
\def\cO{{\cal O}}
\def\cP{{\cal P}}
\def\cS{{\cal S}}
\def\gA{{\goth A}}
\def\x{\bowtie}
\def\ssa{{\em sub specie aeternitatis \/}}
\def\Tr{{\rm Tr}}
\begin{document}
\title{The Space-time Origin of Quantum Mechanics: 
 Covering Law}
\date{\today}
\author{George Svetlichny\thanks{
Departamento de Matem\protect\'atica,
Pontif\protect\'{\protect\i}cia Universidade Cat\protect\'olica,
Rio de Janeiro, Brazil \newline e-mail: svetlich@mat.puc-rio.br}}
\maketitle
\begin{abstract}
Lorentz covariance imposed upon a quantum
logic of local propositions for which all observers can  consistently
 maintain state collapse
descriptions, implies a condition
on space-like separated propositions that if imposed on generally
commuting ones would lead to the covering law, and hence to a hilbert-space
model for the logic. Such a
generalization can be argued if state preparation can be 
conditioned to space-like separated events using EPR-type correlations. 
This suggests that the covering law is related to space-time structure,
though a final understanding of it, through a self-consistency requirement,
will probably require quantum space-time.  
\end{abstract}

\section{Introduction}

The origin of hilbert-space quantum theory has been a
nagging question ever since its creation.
Axiomatic approaches,  by which one
attempts to derive the hilbert-space formalism from postulates
whose content is supposed to be clear  and whose truth is supposed to
be compelling, have only had limited
success. Even if progressive clarity has been achieved, 
 the truth of the axioms never seems compelling. 
Something is missing, and the formalism
continues to mystify. We shall here attempt to dispel  part of
this mystery by arguing that space-time considerations provide
motivation for adopting some of the axioms that are
hard to justify otherwise.

Though there are many axiomatizations of hilbert-space quantum
mechanics, we shall here focus on one,
the well-known and much investigated Piron's \cite{piron}``quantum logic". 
The main object of
consideration is a complete atomic orthomodular lattice of ``physical
propositions". To have a generalized hilbert-space model one has to
assume, among others, an axiom called the ``covering law."  
It is this 
law that has received considerable attention, being
the most controversial of the ingredients.

There are many attempts to reduce the covering law to clearer
and more compelling physical statements, generally by introducing further
structures into the quantum logic, such as measurements, transition
probabilities, propensities, etc.
We show here that some such structures provide us
with  means of deriving necessary conditions on the quantum
logic if it is to describe a Lorentz-covariant
theory. Generalizations of such conditions are then seen to be
sufficient to derive the covering law and thereby a Piron-type
hilbert-space model.

There are three main ingredients in our argument. The first is the
existence of Heisenberg-like physical states that suffer 
``collapse"-type transformations upon measurements. The second is
Lorentz covariance, which, beyond the usual group-action type formulation,
 includes also
what we call ``covariance
of objectivity'' (Postulates \ref{pos:oi} and \ref{pos:oii}) 
of 
section \ref{sec:stm}. 
These state roughly that if a state is
prepared by a measuring apparatus with space-like separated
parts then it has the usual covariance properties with respect to local
observables in regions that are future time-like to all the parts of the
measuring apparatus. An immediate consequence of this assumption is a
condition on space-like separated propositions which, if applied to any
commuting ones, would imply the covering law in existing axiomatic
schemes. This suggests that the whole covering law may have a space-time
origin. To reach such a conclusion however, one has to somehow 
relate time-like
and space-like situations, which leads to the third major ingredient,  
that there are 
sufficiently many states with EPR-type correlations to be able to
prepare arbitrary states conditioned to space-like separated events, as
is the case for ordinary relativistic quantum mechanics. 

Our approach is also of an axiomatic character, and so too suffers
from  the shortcomings we attribute to all such attempts. 
To its merit, it does clarify
the nature of the final
physical basis behind quantum mechanics.  In particular, the third  
 assumption suggests that a final justification could only come
through some form of generalized quantum gravity
where the light cone is not a fundamental but an emergent and object.

We have already argued for a space-time origin of the quantum
formalism 
(Svetlichny \cite{svet3}). There we use the
hypothesis that it is impossible to communicate superluminally
(ISC).
Now the use of ISC to deduce constraints on physical theories must
be considered at best heuristic, for ISC must be traceable to more
basic considerations. In fact, in theories such as quantum gravity,
where the light cone is an emergent  object, ISC
itself must be emergent.
It is thus imperative that we try to
re-establish the putative connection between hilbert space and
lorentzian space-time in a way that makes no appeal to signals. 
To this end we must set up some of the machinery of what could
be called {\em relativistic quantum logic},  quantum logic
subject to the requirements of special relativity. 

Relativistic quantum logic is relatively new, about
two
decades old.  The paper of Mittelstaedt \cite{mitt1} could be said to be one
of the first pioneering works published on the subject. 
The formalism
was applied to the analysis of the Einstein-Podolsky-Rosen experiment by
Mittelstaedt \cite{mitt2} and by Mittelstaedt and Stachow
\cite{mitt3}. The approach
is based on the dialogical (dialog logic) view of physical propositions
upon which relativistic restrictions are applied in the form of
spatio-temporal validity regions. The work of Neumann and Werner
\cite{neumann}
is an elaboration of Ludwig's \cite{ludwig} measurement axiomatics. A 
causality
postulate is introduced for systems prepared in a space-time region and
recorded in a space-like separated region. Examples are presented but no
consequences are derived. The author's  own first ideas were also
developing around the same time, but in contrast to the above mentioned
works were inspired mainly by local algebraic quantum field theory
introduced originally by Haag and Kastler (see Haag \cite{haag}). 
It is this
viewpoint that we follow in this paper. One should also mention the
work of Mugur-Sch\"achter \cite{mugur1,mugur2} which though not
explicitly ``relativistic'' is undeniably spatio-temporal and thus
related.

Though the  idea of relativistic quantum logic is
not new, the application that we have in mind, to seek a
space-time basis for the covering law, {\em is} new. 
It seems that to do so, we must incorporate
structures that go beyond the usual ortho-algebraic ones. 
As a guide we try to
adhere as much as possible to notions current within the usual
``Copenhagen" interpretation and accepted by the greater part of the
physics community.  In doing so, we do not
advocate this interpretation nor claim that
it is in some sense correct, only that it provides a set of principles
that are sufficiently characteristic of quantum mechanics to be an
interesting and familiar starting point. 

 We thus assume the usual notion of {\em ensemble\/}.
Ensembles of physical systems give rise to representational elements in
some abstract set of ``physical states", which for conventional quantum
mechanics is the set of {\em density matrices\/}, positive hilbert space
trace-class operators of trace one.
Ensembles may consist of {\em subensembles \/} in which case these form
well defined fractions given by a real number in \([0,1]\). 
Ensembles and subensembles are to be considered as
potential ontological entities capable of partial realization. Generally
ensembles are partially realized by repetition of preparation procedures
and subensembles identified by the occurrence of some physical results
during preparation. The subensemble fraction is assumed to be
approximated by the frequency of occurrence of the corresponding result.
Besides ensembles of physical systems we can consider ensembles of
measurements or experiments. These are partially realized by carrying
out the corresponding acts a large number of times in such a way that
they do not interfere with each other nor are interfered with by other
acts and events in the universe. For this to make sense one must assume
that one can individuate the necessary physical systems and the
experimental apparatus in a way that warrants neglect of external
influences, and posit some type of relativity theory by which act
performed in different regions of space-time may be considered as
performing the same experiment. Also for each one of these
experiments it is usual to use a mathematical model in which the
corresponding experiment is the only thing existing in all of
space-time. This is a deliberate idealization of the isolation of the
experiment from external influences. We shall tacitly subscribe to all
such usual idealizations and conventions which underlie an ``ensemble"
interpretation of a physical theory.

In the recently introduced ``consistent histories" approach to quantum
mechanics \cite{hartle,omnes}, many of the usual assumptions about
``physical states" become considerably weakened, especially concerning
the ``collapse" of the state due to measurements. A relativistic quantum
logic based on a consistent histories viewpoint would proceed in
a radically different direction, and at first sight would not lead to the
same conclusions, and so provide no justification for the covering law.
We have suggested elsewhere \cite{bialo} 
that quantum gravity would renormalize any such
theory to one in which the covering law holds, as this would be a
fixed point in a self-consistency requirement, however the argument used 
is still
rather sketchy, and so will not be considered here, except for a few
remarks toward the end. Based on this
however, we feel that the present ``collapse-biased" considerations are
pertinent to a final explanation and must be taken into account.

We must call attention to the distinction between ultimate physical
facts and physical descriptions. Physical facts include at least such
uncontroversial   happenings  as  counter   clicks,   collisions,
supernovas, etc., about which all observers agree. Descriptions are
formal tools needed to deal with facts. The distinction is not at all
clear-cut for one generally tries to include among the facts inferred
objects such as the earth's interior, and these may be argued by others
to be just descriptive  constructs that coordinate the true
uncontroversial facts. One may  maintain  that elementary particles are
just formal objects we have invented to provide a more visualizable
description of the surprisingly complex and  subtle antics of
macroscopic bodies. In Feynman-Wheeler electrodynamics there are no
electromagnetic fields, only charged bodies interacting along light-like
intervals. If such a theory is taken as true, then the usual
electromagnetic  fields become just remote and formal descriptive
paraphrases of the facts. A physical theory must make some declarations
as to what is factual and what is descriptive, it must make some
``ontological" commitment, though part of one category may slide over to
the other as one changes the postulated  relation of the descriptive
elements to what are considered ultimate facts. Our concern in this
paper is with classes  of theories in which certain {\em descriptions}
can be consistently maintained regardless of their relation to true
ultimate facts. Descriptions belong to observers and are often
frame-dependent. Ultimate facts are self-subsisting and have nothing to
do with frames. All observers must agree upon them. Relativistic
theories relate ultimate facts placing them in equivalence classes under
the action of an appropriate relativity group. Frame-dependent
descriptions and group action must coexist. This places constraints on
the possible theories. It is some of these constraints that we try to
explicit.

Our exposition, though roughly of an axiomatic nature, will gloss over
mathematical details of purely technical type so as not to overburden
the principal conceptual structure, whose presentation is the aim of
this paper. We do not claim that we've found compelling reasons for
quantum mechanics to be the way it is. We do claim that we've found
a set of physical assumptions that 
can guide a more physically motivated 
axiomatics.  That such a set of
assumptions exists is worthy of note even if some of them can be
seriously questioned in isolation.
 Our results must be considered in any attempt at
unification of space-time with quantum mechanics.

\section{Projection Rule and Objective Mixtures}

Let a quantum state be represented by a density matrix \(\rho\)
and perform an ideal measurement  represented by a self-adjoint operator
\(A\), which for simplicity's sake we assume has a discrete spectrum. 
Thus \(A = \sum \lambda P_\lambda\), where the
sum is  over  distinct eigenvalues \(\lambda\), and the
\(P_\lambda\) are spectral projectors.
The projection rule  states that outcome \(\lambda\) occurs with
frequency \(\Tr(\rho P_\lambda)\) and if this not zero, then the  state
after  the measurement is given by
\(\rho_\lambda = P_\lambda
\rho P_\lambda / \Tr(\rho P_\lambda)\).  We have here a ``beam-splitter"
interpretation of the measurement: the resulting states \(\rho_\lambda\)
for different values of \(\lambda\) are maintained separate either
formally (by conditioning further measurements or even data analysis to
particular outcomes of the current one), or even physically by guiding
the resultant states into different spatially separated regions.  One
can however disregard which outcome occurs and consider each instance of
any of the post-measurement states
 as being an instance of a single state which
would now be represented by \(\rho_A = \sum_\lambda {\rm
Tr}(\rho P_\lambda) \rho_\lambda = \sum_\lambda P_\lambda \rho
P_\lambda\).   This  is usually referred to  as  an  ``incoherent
mixture" of the resulting states \(\rho_\lambda\). One cannot reverse
this, just from the density matrix \(\rho_A\), there is no way of
determining the constituent components \(\rho_\lambda\) and the
corresponding frequencies \(\Tr(\rho P_\lambda)\). Even if we seek pure
components, a non-extreme density matrix \(\rho\) can be decomposed in an
infinite number of ways into a convex combination of extreme matrices,
that is,  there are
infinitely many Borel probability measures \(\mu\) with support in the
subset \(\cP\) of extreme points such that \(\rho = \int p\, d\mu (p)\).
This of course
raises a much ventilated controversy: given a non-extreme density matrix
\(\rho\), is any among its infinite integral representations as a convex
combination of extreme points somehow better, or even objectively or
ontologically ``correct"?  If some principle is assumed by which a
unique representation is singled out, we shall call this representation
an {\em objective mixture}, and to distinguish it from a purely
mathematical integral representation, we shall use an indexed equality
sign \(=_o\) for the former. Thus \(\rho =_o \int p \, d\mu(p)\) 
means that
it is {\em this} representation that is singled out by the postulated
principle. We use the term ``objective" to give a deliberate bias to the
notion, as we envisage that such unique representations have their roots
in some objective reality. In particular, 
prior measurements may provide a basis for such
representations.

One often sees another type of pure-to-mixed state transformation, 
the partial trace. For any density matrix \(\rho\) defined in a tensor
product hilbert space \(H_1 \otimes H_2\) one can 
define the {\em partial
trace\/} \(\rho^{(1)}=\Tr_{H_2}\rho\), a density matrix in \(H_1\),
defined by requiring that for any bounded operator \(A\) in \(H_1\) one
has \(\Tr(\rho^{(1)}A)=  \Tr(\rho (A \otimes I_{H_2}))\).
In general even if \(\rho\) is a pure state,
the partial trace \(\rho^{(1)}_{\psi}\) is not.
What is
usually said about this last situation is that \(\rho^{(1)}_{\psi}\)
represents an ensemble of first members, in an {\em undetermined\/} state,
of a pure ensemble of a two-member composite system. Thus if we write
\(\psi = \sum_i \alpha_i \otimes \beta_i\) where \(\alpha_i \in H_1\) and
\(\beta_i \in H_2\) then one can conceive \(\psi\) as representing a
composite system with two components, the states of one of which are
represented in \(H_1\) and of the other in \(H_2\). If one now makes a
measurement represented by the self adjoint operator \(A\)  only on the
first component, the expected value is \(\Tr(\rho^{(1)}_{\psi}A)\) and so
as far as the measurements on the {\em first} component are concerned,
the system acts as though the first component is in a mixed stated given by
\(\rho^{(1)}_{\psi}\). The conventional wisdom concerning this situation
is however that this is always a mathematical description and no
objective mixture \(\rho^{(1)}_{\psi}\) of any kind is present 
(except for the very particular
case of the partial trace being extreme). The
``partial trace"  state is considered to be ontologically different from
the other types of ensembles. The prototypical example of this situation
is the singlet state of a two photon 
system. If no measurement is made on one of
the photons, then the other one, in so far as it could be construed as a
separate entity, is considered to be ``unpolarized", that is, in no
definite state of polarization. In fact, 
if any polarizer is placed in front of it, the probability
is always one half that it will pass through. One must point out that to
be ``unpolarized" is {\em not\/} a possible state that a photon may be
in, as any one-photon state is always in some state of polarization. Thus
to talk about an ``unpolarized" photon is to employ a (useful) 
metaphor concerning
the presence of an entangled state involving several photons. 

A rather strong principle that leads to objective mixtures 
could be called ``primacy of pure states": given a mixed state, then any
given physical instance of such a state is in fact a physical instance
of a unique pure state which possibly varies from instance to
instance. Primacy is given to pure states and mixed states arise
through mere ensemble mixtures and do not represent new irreducible
ontological entities. This is usually the attitude upheld in elementary
textbooks on quantum mechanics especially for systems comprised of a
small number of particles, as one can easily prepare mixtures that {\it
prima facie\/} seem to obey it. One could seriously question it for
mesoscopic and larger systems. We have argued (Svetlichny
\cite{svet1,svet2})
that its negation can and does lead to interesting possibilities as
there are combinatorial hidden-variable models in which mixed states
violate this principle, opening up a new approach to the distinction
between the classical and the quantum. A explicit revocation of the
primacy of pure states can be found in 
Czachor's \cite{czachor} proposals for non-linear quantum mechanics, in
which 
density matrix evolution is not reducible to evolution
of its component mixtures, the non-uniqueness of which is behind the
causality problems of non-linear deformations of quantum theory.
We shall see below that the
principle cannot be universally upheld along with special relativity and
usual notions of causality. Nevertheless, its simplicity makes it a
useful heuristic device and it does bear examination on two grounds: 1)
it becomes relevant once one contemplates alternative physical theories,
and 2) it's a useful starting point for seeking weaker criteria for
objective mixtures.

Under the primacy of pure states, if  \(\rho =_o \int p \, d\mu(p)\)
and \(\cA\) an observable whose expected value in a pure state \(p\) is
\(<\!\!\cA\!\!>_p\) then the expected value \(<\!\!\cA\!\!>_\rho\) in state 
\(\rho\) has to be
\(\int <\!\!\cA\!\!>_p \, d\mu(p)\). For a conventional quantum mechanical
observable represented by a self adjoint operator \(A\) one 
has \linebreak
\(<\!\!\cA\!\!>_p = \Tr(pA)\) and then \(<\!\!\cA\!\!>_\rho = 
\int \Tr(pA)\, 
d\mu(p) = \Tr\left(\left( \int p \,
d\mu(p)\right)A\right) = {\rm Tr}(\rho A)\) by the 
linearity of the trace and the
operator A. The representation as an objective mixture drops out
and any other representation would lead to the same observable
consequences. Operationally, there is no
observable difference between two different representations, and some
maintain that  objective reality is related to the density matrix itself
and representations as convex combinations of pure states is a purely
mathematical affair. If however  one wants to depart from ordinary
linear quantum mechanics, the question of representation of mixtures
becomes crucial as the integral formula for \(<\!\!\cA\!\!>_\rho\) could 
very well
depend on the representation used in which case some form of objective
mixtures has to be maintained. Furthermore, maintaining some such
version does lead to very interesting and important consequences as one
is then able to influence ``objective reality" at a location space-like
to one's own through EPR-type long-range quantum correlations. 
Consider the singlet state of a
two-photon system when the two photons are space-like separated and,
 say, traveling along opposite arms of an EPR-type apparatus. If we now
perform an observation represented by a non-degenerate quantum
observable \(A\) with normalized eigenvectors \(\psi_1\) and \(\psi_2\)  on
one arm of the apparatus, then by the projection postulate and strict
correlations in the singlet state,  the state on the other arm
immediately after the measurement is an equal mixture of \(\psi_1\) and
\(\psi_2\) represented by the density matrix \({1 \over 2}I = {1 \over
2}(\psi_1, \cdot)\psi_1 + {1 \over 2}(\psi_2, \cdot)\psi_2\). Now, if we
believe in objective mixtures and rewrite this with \(=_o\), we see that
by changing the observable \(A\) to one with a different eigenbasis, we
immediately change the objective mixture on the other arm. This ``action
at a distance" upon supposed objective mixtures has been 
extensively discussed
ever since the original EPR paper (Einstein, Podolsky, and Rosen,
\cite{EPR})
and has recently been used to derive a series of strong constraints on
possible alternatives  conventional quantum mechanics 
(Gisin  \cite{gisin1,gisin2,gisin3,gisin4}; Pearle
\cite{pearle1,pearle2}; Svetlichny \cite{svet3}). Such
constraints stem from the fact that if one is not careful, such
alternative theories will allow for humanly controlled  superluminal
signals  and  hence  supposed 
difficulties  with special relativity. We shall
henceforth refer to these results as the {\em ISC constraints}. 
Theories in which objective mixtures as the result of measurement are 
allowed descriptive elements are strongly constrained and so 
it behooves us to try to give this notion some solid
foundation.

Let us therefore assume that states collapse by  measurements to
objective mixtures and see what this may mean. Consider again a source
of singlet two-photon states which then travel in opposing arms of an
EPR-type apparatus. Assume we are in a reference frame,  
the rest frame,  in  which  the apparatus and the source
is at rest so that the two correlated photons are always at equal
distances along the two arms. Put a vertically oriented linear
polarizer at some distance
along one arm, call it arm {\it 1\/} and mark the other arm, call it arm
{\it 2\/}, at the same distance without first placing anything in the
way of the photon. By the usual arguments we must now conclude that just
beyond the mark one has an objective mixture of vertically and
horizontally polarized photons in equal  proportions.  We shall also
consider  a  frame,  the moving frame, in which
a photon on arm {\it 2\/} reaches the mark before its mate reaches the
linear polarizer. In this frame, just beyond the mark, the photons are
still unpolarized and an objective mixture of linearly polarized photons
comes into being further down the arm. Thus the objective mixture
description is frame dependent. In itself, frame dependence is not a 
defect and we face it all the time.
A static magnetic field, viewed from a moving frame, becomes a
magnetic and an electric field, so the presence or absence of an
electric  field  is frame dependent. However, this  type  of behavior is
easily explained by the notion of covariance under a group
representation while the frame dependence of objective mixtures is of a
completely different nature.
Try now
to give meaning to the statement that in the rest frame, just after the
mark, there is an objective mixture of vertically and horizontally
polarized photons. At first glance one might say that each photon would
pass with certainty either through a vertically or horizontally oriented
linear polarizer and that this is not true for any other ``filters" that
can be placed in its path (uniqueness of objective mixtures). Let us
call this the {\em passage criterion}. Note that in this situation the
criterion is counterfactual for we have no way of knowing which
polarization any individual photon has, but if we {\em did} know, it
{\em would} pass through an appropriately
oriented polarizer. 
Now call upon Maxwell's demon's cousin, the quantum demon. This being
has knowledge of quantum mechanical systems that cannot be achieved by
any humanly constructed apparatus, in our case the knowledge missing in
the counterfactual criterion. Place now a linear polarizer just after
the mark and have the quantum demon rotate it through a sequence of
horizontal and vertical orientations in such a manner as to pass all the
photons from the objective mixture that impinges upon it. How does this
situation look from the moving frame? In this case the demon is twirling
his polarizer in front of {\em unpolarized} photons and even so he is
capable of letting all of them pass through. What's responsible for this
strange fact? We can of course try to blame the linear polarizer that
sits at arm {\it 1\/} but the event of a photon impinging there is to
the future of its mate impinging upon the demon's polarizer. This looks
like inverted causal order, but it's inverted order at space-like
separation and so could be deemed innocuous (though there is the danger
that concatenating two such could lead to time-like retrograde
causality).
Also it's not surprising
that such inverted causal order appears, as we have already posited
something like ``action at a distance" for manipulating distant
objective mixtures and a Lorentz transformation can turn this into
action into the past. What is often desired  of  theories that show such
apparent causal  anomalies  (such  as tachyon theories)  is  that  the 
causal
order  of  events  can  be  {\em reinterpreted} as again to follow a
strict temporal order. This would make the notion of cause and effect
frame dependent, but in the end the usual notions of causality can be
maintained in any frame. Let us call this desideratum upon physical
theories ``strict temporal causality". If we assume this then we cannot
blame the polarizer on arm {\it 1\/} and we must assume that even a beam
of unpolarized photons is an objective mixture of vertically and
horizontally polarized photons. However we can restart the whole
argument now with a circular polarizer on arm {\it 1\/} and conclude
that a beam of unpolarized photons is an objective mixture of left and
right circularly polarized photons. Similarly for any other type of
polarizer. This contradicts the whole idea of objective mixtures as
uniqueness is important, and we must state that the conjunction of
special relativity, strict temporal causality, and the counterfactual
passage criterion for objective mixtures is contradictory.  Note
however that primacy of pure states implies the passage criterion
(counterfactual or not) and so the notion of primacy must be abandoned
as a universal principle if we want to keep the other two.
One way to weaken the passage criterion (or the primacy of pure states)
is to abandon the uniqueness requirement. This is not a desirable step
for us as this would undermine the ISC constraints. However such a step
does bring its insights. We would then admit that an unpolarized photon
already has a well defined value for any of its possible polarizations.
We are now faced with a hidden-variable theory where each relevant quantum
observable has a well defined value. We know that any such theory to be
successful must be contextual and non-local (Redhead \cite{redhead}). It is 
the
appearance of non-locality in this context that is indicative. Primacy
of pure states and the passage criterion seem at first to be local in
nature, however let us examine them from an operational point of view.
Confronted with a given physical instance of a mixed state we call upon
our quantum demon to tell us which particular instance of a pure state,
represented by a normalized vector \(\psi\), we are dealing with. How can
we be sure that the demon tells the truth? We test the state with the
question represented by the hermitian projector \(P_\psi\) upon the
one-dimensional space spanned by \(\psi\). If the test fails, the demon
lied, if it passes we're still not sure but this is the best we can do.
If after a very long run all tests pass, we have strong statistical
evidence to believe the demon is truthful. Now comes the crux of the
matter: in any relativistic field theory, \(P_\psi\) is {\em not} a local
observable. We shall treat this some paragraphs below, but given this, we
see that {\em the notion of objective mixtures cannot be operationally a
local notion\/}, and it is the contradictory attempts to treat it as such
that leads to major difficulties.

On the other hand, in the rest frame, the photons just beyond the mark
seem to behave exactly like an objective mixture of linearly polarized
photons because if we place a linear polarizer in a horizontal
orientation just beyond the mark,  we get exact coincidence with what
happens at the other polarizer.
In the moving frame the roles of the arms are reversed
to maintain strict temporal causality but the same description applies.
Another argument for objective mixtures is that we can reproduce the
quantum demon's exploit if we delay the photon on arm {\it 2\/}
sufficiently. At some point before the mark on arm {\it 2\/}
place a mirror that reflects the
photon to a second distant mirror which then reflects it back to the arm
at a position further down the arm from the first mirror but still before 
the mark,
and at this position place a third mirror that redirects the photon
again outward along the arm. Let the distant mirror be so removed
that information about what happened at the polarizer at arm
{\it 1\/} has time to reach an observer stationed at the mark before the
detoured photon reaches him. The observer then uses this information to
rotate a linear polarizer just beyond the mark so as to allow all the
photons to pass through. Of course now the event of the photon passing
through  the  polarizer on arm {\it 2\/} is inside the  future  light
cone of the event of its mate passing through the polarizer on arm  {\it
1\/} and this then remains true in all frames and we cannot invoke the
argument presented above
which knocked down the counterfactual passage criterion.
So it seems we have all
reason to believe in objective mixtures in this situation. In fact, the
passage criterion is now factual and shows indeed that one deals with an
objective mixture of linearly polarized photons. What happens then as
the photon takes its detour. Does its ontological status changes from
unpolarized to polarized somewhere along its path? If so, when does this
happen? If we give a negative answer to the first question then the
photons have been polarized all along including at points of space-like
separation and by our previous argument we fall back into a particular
hidden variable theory which we are trying to avoid, so the
answer must be positive. A natural answer then to the second question would
be that the change happens when the photon reaches a point along its
path that is light-like to the event of its mate encountering the
polarizer at arm {\it 1\/}. But now this means that a fundamental
quantum mechanical feature changes merely due to a spatio-temporal
arrangement. At space-like separation, given our bias for strict
temporal   causality,   we  can  never  maintain   the (counterfactual)
passage criterion of objective mixtures, but as soon as it becomes
light-like, the passage criterion is applicable (and factual) and shows
that one has an objective mixture of linearly polarized  photons.  This
in  itself  demonstrates  that  quantum mechanics is linked to
space-time structure for the photon becomes polarized just by
penetrating a certain light cone. In other words, {\em besides the 
measuring
apparatus, space-time itself participates in state collapse\/}.

This is in stark contrast with Galileian covariant theories. In such
theories the primacy of pure states can be maintained universally
and hence the passage criterion for objective mixtures can be used
in all circumstances.

Now we must look again at the original space-like situation. We saw
we cannot use the counterfactual passage criterion yet we would like
to maintain some version of objective mixtures. The answer that
offers itself is that whereas for time-like situations objective
mixtures are ultimate physical facts, for space-like separations
they are physical descriptions. By this we do not mean they are
arbitrary or  ``unreal" but that the criteria for their choice can
depend on specifics of experimental arrangements and inertial
frames. Thus we should not be surprised that if we place
 a circular polarizer
placed just beyond the mark, in the rest frame there are (according
to the {\em description}) linearly polarized photons impinging on
it, whereas in the moving frame (again according to the {\em
description})
there are no linearly polarized photons anywhere near it.
We can base our criterion for objective mixtures on a
previous measurement event, according to the time order in the given frame. 
Thus in the rest frame it is the linear
polarizer on arm {\it 1\/} that determines the objective mixtures in the 
time
interval between one photon having reached its polarizer and its mate
the other polarizer, and in the moving frame it is the circular
polarizer on arm {\it 2\/}. It is also part of the conventional wisdom that
this is a consistent way of proceeding and that observers in either
frame, each one using his own description, will agree as to their
predictions about ultimate physical facts.

Relativity theory forces us to abandon a naive picture of
primacy of pure states and to adopt a sort of hybrid view in which
objective mixtures are ultimate physical facts in some situations
and physical descriptions in others. In a fixed frame one
situation blends into the other without apparent discontinuity as
soon as certain spatio-temporal relations are achieved.

The paradoxical nature of objectifying too much the state description
after measurement in relativistic theories was also pointed out by Mielnik
\cite{mielnik} who concludes that state-reduction and relativity are
mutually inconsistent. This is a surprising conclusion as relativistic
quantum field theory is highly successful. He is led
to this by however by tacitly admitting results of counterfactual
experiments. Applied to the singlet two-photon state
considered above, his reasoning would lead to the same hidden-variable 
theory that the
counterfactual passage criterion does. Disallowing such counterfactual
definiteness blocks the contradiction. A different resolution of
Mielnik's paradox is given by Finkelstein \cite{fink}.

Return again to the projection postulate and let \(A\) be a
self adjoint operator with discrete spectrum and spectral
decomposition \(A = \sum \lambda P_\lambda\). Let the normalized vector
\(\psi\) represent 
the state
upon which the observation is performed, and let \(\psi_\lambda =
P_\lambda \psi / ||P_\lambda \psi||\) whenever \(||P_\lambda
\psi|| \neq 0\). We now write \(\rho_A =_o \sum ||P_\lambda \psi||^2
(\psi_\lambda,\cdot)\psi_\lambda\)  to  indicate  that  \(\rho_A\)  occurred  
through  a
previous
preparation (measurement) in which, had the ``beam splitter"
viewpoint been adopted, the \(\psi_\lambda\) would describe the pure
state in each ``beam".  Objective mixtures are then tokens of
preparations. Whether such a mixture is actual and complies with the
passage criterion or descriptive,  depends now on some
spatio-temporal situation.  In all cases however one has a
{\em correlation } criterion. If immediately after observing \(A\) we
perform a test of whether the resulting state is \(\psi_\lambda\) 
(by using the orthogonal projector
onto this vector as the observable), then
the test is satisfied if and only if \(\lambda\)
occurs (exact correlation). This of course is due to the orthogonality of the
\(\psi_\lambda\) for different \(\lambda\). This is not true for any other 
set of
one-dimensional projectors associated to those outcomes \(\lambda\) for
which \(P_\lambda \psi \neq 0\). Thus the correlation
criterion picks out a unique convex combination of pure states and
so is a legitimate basis for objective mixtures. Being an
operational criterion (at least for ensembles)
it is about as ``objective"
as one can wish. It works both in space-like and time-like
situations. Its major disadvantage is that it doesn't apply to the
state in itself but involves the preparation that produced the
state. Such knowledge of the preparation procedure can be used to
separate the beams again if the original measurement was not of the
beam splitter type. Just measure \(A\) again  (or any other observable
for which the \(\psi_\lambda\) are eigenstates) and adopt the beam
splitter attitude.

For future reference, let us examine now what happens when we perform a 
second measurement
represented  by  a self-adjoint operator $B$ with  discrete
spectrum
and spectral decomposition $B = \sum \mu Q_\mu$. We now have $(\rho_A)_B
= \sum_\mu Q_\mu \rho_A Q_\mu = \sum_{\mu \lambda} (Q_\mu
P_\lambda\psi, \cdot)Q_\mu P_\lambda\psi$. Now can this be interpreted as
an {\sl objective} mixture according to the correlation criterion?
If by this we mean that each distinct (unnormalized) final state $Q_\mu
P_\lambda\psi$ is uniquely correlated to a set of outcomes, then yes. If
however we want that there be a test for each final state which passes
if and only if that state is produced, or if we want to ``separate the
beams'' as was done for a single measurement above,
then the distinct final states
must be orthogonal. We
shall consider the objective mixture description legitimate in this
case. This latter situation is always
realized for any initial state $\psi$ whenever $A$ and $B$ commute.

How should one interpret then the ISC constraints? They must now be read
in the following manner: in any theory for which the objective mixture
criterion is the presence of an adequate previous measurement and for
which such  {\em descriptions} form a consistent logical system leading
to frame independent predictions of ultimate physical facts, the results
of the references hold.  So reinterpreted, the validity of the works is
maintained and they still provide strong criteria for selecting
theories, but these theories are to be chosen among special types.
This is an important insight, as one can now see under what conditions
ISC may not constrain a theory or constrain it less. Thus if ``state
collapse" as a descriptive element, along with ISC, argues strongly for
a linear theory (as is expounded in \cite{gisin2,svet3}), theories
without this descriptive element, such as the consistent-histories
approach to quantum mechanics, may possibly be made non-linear without
violating ISC. We argue in this direction in \cite{svetlichny:quantum, 
svet5} and
take up this point later in this paper.

\section{Locality and Purity}

Since the idea of objective mixtures involves decomposition into
pure states, we must examine the nature of pure states in
relativistic quantum mechanics. There are roughly two
approaches to this, via fields (Streater
and Wightman \cite{streater}) and via
algebras of observables (Haag \cite{haag}).

The algebraic approach is closer in spirit
to what we are contemplating here as it introduces {\em algebras of
local observables}, that is, it associates to each bounded region \(\cO\) of
space-time an algebra \(\gA(\cO)\) of observables that correspond to
experiments that can be executed in \(\cO\). In relativistic quantum logic
we would be interested in propositions that can be tested
in \(\cO\) and so should in some natural way be related to the algebra
\(\gA(\cO)\). It is
however rather difficult to come across examples of local algebras
except through quantum fields and so we present here one possible
construction. Assume for simplicity that we have a real (uncharged)
scalar relativistic Wightman field \(\Phi\).
Such a field is an
operator-valued distribution in the sense that there is a fixed dense
domain \(\cD\) such for any \(f \in \cS({\bbf R}^4)\) there is an 
essentially
self-adjoint
operator \(\Phi(f)\) on the invariant domain \(\cD\) and such that
for all \(\phi, \psi \in \cD\) the map
\(f \mapsto (\phi,\Phi(f)\psi)\) defines a tempered distribution.
Consider now the operators \(\Phi(f)\)
for \({\rm supp}\,f \subset \cO\) for some bounded region of space-time 
\(\cO\). Let
\(\gA^c(\cO)\) be the set of bounded operators \(A\) such that \(A\cD 
\subset
\cD\) and which commute on \({\cal D}\) with all the operators \(\Phi(f)\)
introduce above. Define
the von-Neumann algebra \(\gA(\cO)\) as \((\gA^c(\cO))'\), the
commutant of \(\gA^c(\cO)\). The algebra \(\gA(\cO)\) is then taken to be
the algebra of observables in \(\cO\). An alternative definition would be to 
take for
\(\gA(\cO)\) the von-Neumann algebra generated by the bounded functions of
the \(\Phi(f)\), that is by those operators of the form \(F(\Phi(f))\) where
\(F\) is a real Borel-measurable bounded function on \({\bbf R}\).
The hermitian projectors in \(\gA(\cO)\) should then correspond to the
testable proposition in \(\cO\). Now it is a know fact (Araki
\cite{araki}, Haag
\cite{haag}) that the von Neumann algebra \(\gA(\cO)\) is of type III. This 
means
that it contains no finite-dimensional projector and in particular no
one-dimensional projector. One-dimensional projectors are indicator
propositions for pure states, that is if \(P = (\psi,\cdot)\psi\) 
for a normalized vector \(\psi\), and
\(\rho\)  is a density matrix, then one has \(\Tr(\rho P) = 1\) if and only
if \(\rho = (\psi,\cdot)\psi\). It is these projectors that must be used
to test for pure states, and therefore  purity of states is not a local
notion. This is the basic insight that relativistic quantum field theory
provides for quantum logic. A consequence of this  is that the notion of
objective mixtures becomes a non-local notion, and in particular the
correlation criterion, as it involves testing for pure states, is a
non-local criterion. Now if one cannot test for purity by projectors in
\(\gA(\cO)\) for a bounded region \(\cO\), it is also highly plausible that
one can test for pure states by projectors in the algebra associated to
any {\em time slice}: \(\cO = \{(x,y,z,t)| t_1 < t < t_2\}\). Such an
algebra is defined by an appropriate limiting procedure in term of the
algebras associated to bounded regions contained in the time slice.
For free
fields any time-slice algebra is just \(B(H)\), the full operator algebra
of the physical hilbert space. In general one expects the field to obey,
in some appropriate sense, a hyperbolic differential equation and so the
field values at any point can be determined from their values in a time
slice. This would mean that the time-slice algebra coincides with the
algebra associated to all of space-time, which again should be \(B(H)\)
for reasonable theories.

Suppose now that we perform a measurement in a bounded space-time
region \(\cO\) upon a pure Heisenberg state represented by a normalized 
vector
\(\psi\).
Consider at a space-like point to \(\cO\) two observers, one
for whom, according to the time variable in his frame, the
measurement is still to happen and another one for whom the
measurement has already taken place. As they fly by each other they
exchange notes, each one indicating what the quantum state is. One
says it's pure, the other one mixed. This apparent contradiction
disappears when one realizes that an  operational definition of
purity  is  not local. Each one's assessment depends on  a  time-slice
which for one is prior to the experiment and
posterior for the other. Thus in relativistic quantum mechanics, in
the presence of measurements, whether a state is pure or not is
frame dependent. If both observers are however in the future light
cone of all points of \(\cO\), then their assessments of the state
agree, both will say it is mixed and their respective descriptions
should differ merely by the action of a representation of the
Lorentz group, that is by usual Lorentz covariance. There is a point
of consistency here: in the region where the observers do not
agree, no local observable should be able to distinguish between the
two conflicting descriptions.  This in fact is the case, for let the
experiment be
represented by a self-adjoint operator \(A\) with discrete spectral
decomposition \(A=\sum \lambda P_\lambda\) and let \(B\) be a
self-adjoint operator pertaining to a space-like
separated local algebra. Let us say the first observer's description
is that \(B\) is being observed on \(\psi\). The expected value then
would be \((\psi, B\psi)\). Assuming normal Lorentz covariance, the
second observer describes the situation as observing \(U(g)^*BU(g)\)
upon \(U(g)^*\rho_AU(g)\) where \(U(g)\) is a unitary operator
representing the element \(g\) of the Lorentz group that connects the
two observers. This last description gives a mean value of \(
\Tr(U(g)^*BU(g)U(g)^*\rho_AU(g)) = \Tr(B\rho_A) = \Tr(B \sum
P_\lambda P_\psi P_\lambda)\)
where \(P_\psi = (\psi, \cdot)\psi\). By the
linearity and permutation symmetry of the trace this is \(\sum
\Tr(P_\lambda B P_\lambda P_\psi)\). Now in all relativistic
field theories, local observables in space-like separated regions
commute, so one has \(P_\lambda B P_\lambda = P_\lambda^2 B =
P_\lambda B\) and since \(\sum P_\lambda = I\), the expected value is
\(\Tr(BP_\psi) = (\psi, B \psi)\) exactly as for the first
observer. Hence the discrepancy in description of purity of states,
due to the non-local nature of the correlation criterion, has no
effect on locally observable quantities and this is precisely what
accounts for the consistency of this criterion.

We are now in position to abstract from the above situation in
conventional quantum mechanics and introduce a sketch of an axiom system
for a measurement theory in relativistic quantum logic.

\section{Propositions, Properties, States, and Ensembles}

The basic ingredient of a ``quantum logic" approach to a physical theory
is a pair \((\cL, \cS )\) where \(\cL \) is an orthomodular poset
(usually a lattice) of physical {\em propositions\/}, and \(\cS \) an
abstract convex set whose elements correspond to physical {\em
states\/}. The set \(\cP \) of extreme points of \(\cS\) correspond to
{\em pure states\/}. The relation between \(\cL\) and \(\cS\) is usually
that elements of \(\cS\) are \(\sigma\)-additive probability measures on
\(\cL\), and given \(s \in \cS\) and \(a \in \cL\) the number \(s(a)\)
corresponds to the {\em probability\/} that the proposition \(a\) be
true in the state \(s\). We shall not necessarily adhere to such a view
though much of the literature adopts it. We shall not here go into
details about how one operationally prepares pure states nor determines
their purity except in so far as is needed to consider the space-time
relations involved.

One generally sees two concepts of physical state that. The
most common one is that of an {\em instantaneous state \/}, and a
physical system is supposed to at each time instant {\em be\/} in some
instantaneous state. The other notion is that of a state \ssa in which
case the whole temporal history from \(t= -\infty\) to \(t=+\infty\) is
subsumed in the notion. Of course this is an idealization as normally a
state is prepared at some instant and destroyed at some future instant
both by processes foreign to its ``normal" isolated temporal evolution.
Thus to maintain a view \ssa one has to rely on some deterministic
evolution extendible to both temporal infinities. This also means that
the states is considered as a self-subsisting entity. The ability to
extend its evolution to a time prior to its creation, means that it could
have been created at a different time and so it has no knowledge of its
creation, and the ability to extend its evolution to a time beyond its
destruction means that it has no presage of its demise. Such a physical
state is an ontological entity all to itself. The two views coexist in
ordinary quantum mechanics whereby the instantaneous state view is
maintained in the Schr{\"o}dinger picture and the other in the Heisenberg
picture. The \ssa view is more convenient for space-time description as
the notion of instantaneous state, even for normal time evolution,
brings in frame-related considerations due to the frame-dependent nature
of the notion of ``instant". Thus we maintain the \ssa viewpoint in this
paper. Of course even this view cannot entirely avoid frame-related
notions as this type of state is allowed to undergo change through the
measurement process or other external interventions. Such changes are
generally held to be instantaneous in some frame and so if the state is
conceived as having a space-time extent, a frame dependent change of
description is involved. We shall in fact be interested in states having
sufficiently large space-time extents to be able to perform independent
measurements at spatially separated distances.

Just as in quantum mechanics, we shall assume the notion of ensembles
at least in so far as they can be partially realized by repetitions of
preparation procedures. The notions of subensemble and subensemble
fraction is also maintained. Ensembles are represented by elements of
\(\cS\) and we admit, just as in quantum mechanics, that the same element
of \(\cS\) could very well represent many ontologically distinct ensembles
so one should not conflate the two.

\section{Measurements and Objective Mixtures}

We shall assume at least that \(\cL\) and \(\cS\) are related through
measurement. By an {\em instrument\/} \(I\) we shall mean an
exhaustive n-tuple \((a_1,\dots,a_n) \in \cL^n\) of mutually exclusive
propositions; that is, \(a_i \perp a_j\) for \(i \neq j\) and
\(\bigvee_{i=1}^n a_i = 1\). We shall assume that such n-tuples
correspond to physical measurements with \(n\) mutually exclusive and
exhaustive outcomes; that is, when such a measurement procedure is
executed, one and only one of the outcomes occurs. We shall now make
a series of assumptions concerning the act of measurement and
discuss them later.

\begin{postulate}[M1 -- Frequency] \label{pos:m1}
Given an instrument \(I=(a_1,\dots,a_n)\) and a state \(s\), then associated
to an ensemble of measurements of \(I\) in \(s\), is a {\em frequency
function} \(\omega_i^I(s)\) where \(\omega_i^I(s) \geq 0\) and 
\(\sum_{i=1}^n
\omega_i^I(s) = 1\).
\end{postulate}
We've used the term ``frequency function" as a neutral alternative to
``probability" or ``propensity" as there is no need to enter into
interpretational questions at this moment. Of course, the subensembles
of measurements corresponding to the occurrences of distinct \(a_i\) do not
overlap.
\begin{postulate}[M2 -- State transformation] \label{pos:m2}
A state subject to a measurement undergoes a transformation to
another state representable by an element of \(\cS\). Let \(\pi^I : \cS
\rightarrow \cS\) be the map representing this transformation.
\end{postulate}
We are adopting here what in the quantum mechanical case we called the
``incoherent mixture" view of measurement.
\begin{postulate}[M3 -- Subensemble] \label{pos:m3}
Given a state \(s\) the transformed state \(\pi^I s\) consists of
subensembles corresponding to each particular outcome of the
measurement. Each such subensemble is a fraction of the total ensemble
given by the frequency function. Thus there are partial maps \(\pi^I_i\)
such that \(\pi^I s = \sum_{i=1}^n \omega_i^I(s) \pi^I_i s\). The
expression \(\pi^I_is\) is considered to be defined only when
\(\omega_i^I(s) \neq 0\).
\end{postulate}
This assumption allows us now to also adopt the ``beam splitter" view of
measurements.
\begin{postulate}[M4 -- Ideality] \label{pos:m4}
Measurements are ideal, in the sense that for any pure state \(p\) and
any instrument \(I\) one has that \(\pi^I_ip\) is also pure, when
defined.
\end{postulate}
\begin{postulate}[M5 -- Objectivity] \label{pos:m5}
The ensemble produced by a measurement is an objective mixtures of the
subensembles corresponding to the individual outcomes. That is,
\(\pi^I s =_o \sum_{i=1}^n \omega_i^I(s) \pi^I_i s\).
\end{postulate}
In particular, if one performs a measurement on a pure state, then by
M4 and M5 one obtains an objective mixture of pure states.

We leave open as to what exactly is the criterion for objective
mixtures. The precise nature of this criterion is not as
relevant as some of its desirable properties to which we shall draw
attention in due time. We mention, for the sake of concreteness, a
possible correlation-type criterion just as in quantum
mechanics:
\begin{quote}(Correlation Criterion of Objectivity) The criterion
for an observer in the coordinate future to the measurement by an instrument 
\(I\) on a state \(s\) to maintain \(\pi^I s =_o \sum_{i=1}^n \alpha_i s_i\)
with \(\alpha_i \neq 0\) is the availability 
of dicotomic instruments  
\(J^{(j)}=(b^{(j)}_1, b^{(j)}_2)\) such that 
\(\omega_1^{J^{(j)}}(\pi^I_i(s)) = \delta_{ij}\), and
 a strict correlation of the first outcome of
 \(J^{(j)}\) with the \(j\)-th outcome of \(I\), in which case
 \(\alpha_i=\omega^I_i(s)\) and \(s_i=\pi^I_i(s)\). 
\end{quote}
This criterion would suffice for what follows but others could probably
do just as well.

One consequence of assuming the existence of objective mixtures is that
if \(s =_o \int p \, d\mu(p)\) then \(\omega_i^I(s) = \int \omega_i^I(p) \,
d\mu(p)\) and so it's enough to know the frequency function only in pure
states, and we can assume the functions \(\omega_i^I\) are affine.
The same goes for the map \(\pi^I\).

How plausible are these assumptions? Taken together they express our
ability to individuate physical system by appropriate ``filtering"
through measurements and then to perform statistical experiments on
situations so created. This constitutes the basis of normal physical
experimental practice, so to negate this is to radically change our view
of the statistical nature of physical phenomena and our experimental
access to them. Of course these assumptions are already idealized, and
one could argue with the details of each one individually, but some such
set must be postulated to even begin a formulation of a statistical
science. Part of the above assumptions incorporate the notion of 
self-subsisting physical states, that is the \ssa view. This is not a
logically necessary ingredient of a physical theory, ingrained as it may
be. All a physical theory must be able to do is
predict the joint probabilities of events, and the mediation of these by
physical states under evolution is not a necessity. As was mentioned
before, the consistent histories approach does not make use of such a
notion, and the preceding assumptions would have to be modified if one
were to axiomatize such a viewpoint.

\section{Space-time Structure of Measurements}\label{sec:stm}

The relation between a physical theory \((\cL,\cS)\) and space-time
structure generally comes in through external considerations. Ordinary
hilbert-space quantum mechanics admits both Galileian and Lorentz
covariance; such considerations only enter through unitary
representations of an appropriate group and have no expression in the
fundamental formalism as such. In our context we cannot do much better
while some basic mathematical questions have yet to be settled. We
assume lorentzian space-time, and in relation to the physical theory we
assume a series of postulates generalizing some of the usual external
connections already seen in hilbert-space theory.

\begin{postulate}[S1 -- Localization] \label{pos:s1}
Given a bounded space-time region \(\cO\) then there is a
sub-orthomodular-poset \(\cL(\cO) \subset \cL\) corresponding to
propositions testable in \(\cO\).
\end{postulate}
This assumption reflects the notion that experiments are essentially
bounded in space-time.  Usually one also feels that propositions
referring to unbounded regions should only be admitted as
idealizations, that is, limits of local ones. Thus one could postulate
that the union of the subsets  \(\cL(\cO)\) is join-dense in \(\cL\) which
is then viewed as  a set of ``quasi-local" propositions. Quantum field
theory teaches us that one should not assume about \(\cL(\cO)\) properties
that one usually postulates about \(\cL\); thus while the latter is often
taken to be atomic and atomistic, this should not be the case for the
local posets.

We say an instrument \(I = (a_1,\dots,a_n)\) {\em belongs} to a region
\(\cO\) if \(a_i \in \cL(\cO)\) for each \(i\). For two regions \(\cO_1, 
\cO_2\)
we write \(\cO_1 \x \cO_2\) in case they are space-like separated, that
is,
every point of one is space-like to every point of the other. For
\(a_1,a_2 \in \cL\) we write \(a_1 \x a_2\) in case \(a_1 \in \cL(\cO_1)\) and
\(a_2
\in \cL(\cO_2)\) for some regions \(\cO_1 \x \cO_2\). For two instruments
\(I\) and \(J\) we write \(I \x J\) in case they belong to space-like
separated regions.

\begin{postulate}[S2 -- Locality] \label{pos:s2}
If \(\cO_1 \x \cO_2\) then every element of \(\cL(\cO_1)\) commutes with
every element of \(\cL(\cO_2)\).
\end{postulate}
It is customary in the lattice-theoretic approach to equate
commutativity of propositions with their commensurability. It is
likewise customary to assume that space-like separated regions are
causally disjoint (locality or causality assumption). This then is
a major assumption relating space-time structure and the poset of
proposition. We shall write \(a \leftrightarrow b\) whenever \(a\) and \(b\)
commute. It is interesting to mention that in Mittelstaedt's
\cite{mitt1}
scheme, commutativity of space-like separated propositions is necessary
for logical consistency.

If \(I = (a_1,\dots,a_n)\) and \(J = (b_1,\dots,b_m)\), are two instruments
such that \(a_i \leftrightarrow b_j\) for all \(i\) and \(j\), in particular
if \(I \x J\), then we can form a new instrument \(I \wedge J = (a_i \wedge
b_j)_{i=1,\dots,n;j=1,\dots,m}\)

Now the execution of an experiment consists of physical acts leading to
physical results. This in itself has nothing to do with our description
of the experiment nor consequently with the adoption of any particular
reference frame for space-time. The propositions being tested however do
have something to do with reference frames. If an observer finds at time
\(t_0\) that a proposition \(a\) is true about a state \(s\), then the truth
of \(a\) is aspatial, hence to be considered as such at all points of
space. If, as is customary, one considers that a proposition can {\em
become\/} true at a time instant \(t_0\) then it must become true
instantly and simultaneously at all space points. An observer in a
different frame has a different plane of simultaneity so his
propositions become true in a different manner.  This frame dependence
of truth values and becoming-true is not necessarily in conflict with a
frame-independent physics. The mutual consistency of the two however,
given other requirements, can and does lead to constraints on physical
theories.

\begin{postulate}[S3 -- Measurement instant] \label{pos:s3}
If an experiment corresponding to an instrument belonging to a region
\(\cO\) has been executed, then every observer assigns a unique time
instant at which the experiment is considered to be realized and at
which instant one of the propositions related to the instrument  becomes
true and the others false. The plane of simultaneity of this instant
intersects \(\cO\).
\end{postulate}

This assumption seems to be generally, even if grudgingly, accepted. The
experimenter often doesn't have control over the instant in question
which is somehow decided by the physical processes that the experiment
unleashes, yet such an instant is generally identified. Even more often
the case is that the realization of the experiment is associated to some
unique event (counter click, for instance) and the instant is just the
time coordinate of this event in the given frame. In this case different
observers can agree on the same event, which lies in
\(\cO\). There are of course important apparent exceptions to the single
event viewpoint such as coincidence experiments, but even in this case
some maintain that the experiment is only over when all the information
is gathered in some single recording device, such as a brain, in which
case one falls back on the single event hypothesis. We shall not however
adhere to this viewpoint.

Consider now two experiments executed in space-like separated regions.
Let \(I = (a_1,\dots,a_n)\) and \(J = (b_1,\dots,b_m)\) be the corresponding
instruments. Suppose that an observer assigns the same time instant \(t_0\)
to the realization of both experiments. Then, by the commensurability of
the two instruments, at \(t_0\) the observer maintains not only that one of
the \(a_i\) becomes true and that one of the \(b_j\) becomes true but also
that one of the \(a_i \wedge b_j\) becomes true. As far as the observer is
concerned, the two separate realizations of \(I\) and \(J\) is
indistinguishable from a single realization of \(I \wedge J\). This
indistinguishability is our next assumption.

\begin{postulate}[S4 --  Confluence of simultaneous  measurements] 
\label{pos:s4}
If an observer assigns the same time instant to the realization of two
space-like separated experiments, these can be treated equivalently as
the realization of a single experiment whose outcomes are the
conjunctions of the outcomes of the separate experiments.
\end{postulate}
Another observer in a different frame would generally see a time
interval between the two realizations, either \(I\) first, followed by
\(J\), or vice-versa. In this case he must consider the two experiments
as consecutive and for him the becoming-true of propositions related
to the two experiments do not occur simultaneously. If physics is
frame-independent then the two observers must agree about ultimate
physical facts. For this to be the case, the pair \((\cL, \cS)\) must
satisfy certain constraints.

What we still lack are assumptions that express the supposed frame
independence of ultimate physical facts. Just as in hilbert-space
theory, we introduce this via group action.

\begin{postulate}[C -- Lorentz covariance] \label{pos:c}
Let \(\cG\) be the Poincar\'e Group. There are actions \((g,a) \mapsto
\lambda_g a\) of \(\cG\) on \(\cL\) and \((g, s) \mapsto \sigma_g s\) on
\(\cS\) such that:
\begin{enumerate}
\item For all \(g \in \cG\), \(\lambda_g\) is an orthomodular-poset 
isomorphism.
\item \(\lambda_g \cL(\cO) \subset  \cL(g(\cO))\)
\item For all \(g \in \cG\), the action \(s \mapsto \sigma_g s\) is
affine and \(\sigma_g (\cP ) \subset \cP \).
\item Given an instrument \(I = (a_1,\dots,a_n)\), let \(\lambda_g I =
   (\lambda_g a_1,\dots,\lambda_g a_n)\). One then has:
\begin{eqnarray*}
\omega_i^I(s) &=& \omega_i^{\lambda_g I}(\sigma_g s) \\
\pi^{\lambda_g I}_i \sigma_g p &=& \sigma_g \pi^I_i p
\end{eqnarray*}
\end{enumerate}
\end{postulate}
For simplicity's sake we shall write \(g\cdot a\) and \(g\cdot s\) 
instead of
\(\lambda_g a\) and \(\sigma_g s\).

Most of this assumption is a fairly straightforward rendition of rather
standard covariance conditions on physical theories. There are two ways
of understanding the group action. The first, or ``passive" view, is
that if an observer describes an experimental procedure as that of
executing \(I\) upon state \(s\), then another observer whose frame is
obtained from that of the first one by action of \(g\) will describe the
same procedure as that of executing \(g^{-1}\cdot I\) 
upon \(g^{-1}\cdot s\). The
``active" view states that there is another procedure (whose
execution has a clear operational relation to the execution of the first
one) described by \(g\cdot I\) and \(g\cdot s\) 
by the first observer and whose
description by the second observer is by \(I\) and \(s\). The equivalence of
the two views is the essence of relativistic theories.

One possible strengthening of C3 would be to assert that objective
mixtures map to objective mixtures, that is, if \(s =_o \int p \,
d\mu(p)\), then \(g\cdot s =_o \int g\cdot p \, d\mu(p)\). This is
{\it prima facie \/}  a natural assumption, and we shall formulate 
a version of it. However 
we shall also  need  a more subtle manifestation of covariance in relation
to objective mixtures. Since our only assumption concerning objective
mixtures is that they come about through measurements and since the
corresponding ``collapse" is frame dependent, covariance
of
objective mixtures is a rather more involved concept since whatever
criterion for objective mixtures one may adopt, one should not think of
it as a local criterion as the quantum field theory case teaches us. One
must imagine that such a criterion would utilize something like a
time-slice region, idealized say to a space-like hyperplane. In some
regions two observers would disagree even as to which (if any) of the
relevant experiments have been carried out and in which order, leading
to objective mixtures not related by the action of the Poincar\'e group.
Objective mixtures as
observer related descriptions would have varying relation to physical
facts. Thus the simple relation for objective mixtures given in the
beginning of this paragraph can be maintained as referring to the
sate in question only by a pair of observers that are both in the future
light cone of all the relevant measurement processes that enter into the
objective mixture criterion. 

\begin{postulate}[O~I -- Covariance of objectivity I]\label{pos:oi}
If \(s =_o \int p \,
d\mu(p)\), then \(g\cdot s =_o \int g\cdot p \, d\mu(p)\).
The right hand side of these equations refers to the state in question
as seen by two observers related by the element \(g\) of the
Poincar\'e group provided both are in the time-like future of all events
involved in the objective mixture criterion.  
\end{postulate}

Let now \(I =(a_1,\dots,a_n)\) and \(J=(b_1,\dots,b_m)\) be two instruments
and assume \(I \x J\). Consider now an observer, call him Observer 1,
who assigns the same time instant to the realization of both experiments
upon a pure state \(p\) (note well: we do not mean that each experiment
acts on its own ``copy" of \(p\) but each on the {\em same \/} spatially
extended state \(p\)). According to S4 we can view this as a realization
of \(I \wedge J\) and so by the measurement assumptions the state is
transformed into \(s =_o \sum_{i,j}\omega_{i,j}^{I \wedge J}(p)\pi^{I
\wedge J}_{i,j} p\). Another observer, call her Observer 2, will
describe the situation as the realization of \(g\cdot I\) upon \(g\cdot
p\) followed by a realization of \(g\cdot J\) upon the resulting state
of the first measurement. The first experiment produces a state \(s_1
=_o \sum_i \omega_i^{g\cdot I}(g\cdot p)\pi^{g\cdot I}_i g\cdot p\) and
this is transformed by the second experiment into (note the lack of the
objective equality sign) \(s_2 = \sum_{j,i}\omega_j^{g\cdot
J}(\pi^{g\cdot I}_i g\cdot p) \omega_i^{g\cdot I}(g\cdot p) \pi^{g\cdot
J}_j\pi^{g\cdot I}_i g\cdot p\). Our next assumption is that this is in
fact an objective mixture:

\begin{postulate}[O~II -- Covariance of objectivity II]\label{pos:oii}\ \  
\begin{description}
\item {\rm a)} \(s_2 =_o \sum_{j,i}\omega_j^{g\cdot J}(\pi^{g\cdot I}_i
g\cdot p) \omega_i^{g\cdot I}( g\cdot p) \pi^{g\cdot J}_j\pi^{g\cdot
I}_i g\cdot p\).
\item{\rm b)} \(s_2 = g\cdot s\).
\end{description}
\end{postulate}

This is a subtle point. One has no grounds on the basis of previous
assumptions or analyses to claim that the expression for \(s_2\) is an
objective mixture and in fact one needs an additional argument to adopt
this hypothesis (it is true that M5 implies that \(s_2\) is an objective
mixture of certain {\em non-extreme} states, but this is not what we are
saying here, and M5 does not imply O~II). One saw in the quantum mechanical
discussion that two successive measurements with commuting instruments
lead to an objective mixture of the type expressed by \(s_2\) according
to the correlation criterion. One could be tempted to use this as
guideline and adopt a similar hypothesis in the quantum logic case,
seeing that space-like separated instruments commute. This however has
no more compelling reason to be than the covering law itself and we do
not view O~II as a statement about commutativity but about covariance. As
has been pointed out, observers in different frames may even sustain
objective mixture descriptions not related by the action of an element
of the Poincar\'e group. Such disagreements as to the {\em
description\/} of the state, objective as they may be, should of course
not be detectable by local observations just as in the quantum field
theory case, and this leads to definite constraints on the theory, but
these as we shall see will be automatically satisfied. Let us however
concentrate on the set of points that are future time-like to all events
involved in the measurement process. As we found out in the quantum
mechanical case, in this region, the passage criterion for objective
mixtures can be maintained and actually carried out as we can have
knowledge of the measurement results. This means that we can actually
use the ``beam-splitter" view of measurements and consider conditioning
further measurements to the outcomes of the past measurements \(I\) and
\(J\). Now if we condition to outcome \(a_i\) and \(b_j\), then these
outcomes are ultimate facts whether they are simultaneous or not. In the
future region we expect, by the very notion of Lorentz covariance, that
the state defined by this ``beam" should behave in a purely covariant
fashion, that is, obeying C, when tested by local observables. As we
move further and further into the future, local observables in this
region can have increasingly greater space-like extent and
asymptotically can approach time-slice observables. As these are
presumed to be involved in the objective mixture criterion, these are
then also subject to C. In particular, if Observer 1 finds the state to
be \(s_{ij}\) and if Observer 2 is related to Observer 1 by and element
\(g\) of the Poincar\'e group, she should find the state to be
\(g^{-1}\cdot s_{ij}\). Thus we can argue that the given representation for
\(s_2\) should be an objective mixture purely on the grounds of Lorentz
covariance. We should also have \(s_2 = g\cdot s\). This type of
covariance is if course somewhat different from the one expressed in C
as that one makes no reference to measurements. Observer 1 has a
different geometrical relation to the measurements than Observer 2 as in
his situation there is only one plane of simultaneity related to the
measurements, whereas for her there are two. The geometrical structures
are not group transforms of each other. The frame-dependent description
of the measurement process is in contrast with the conventional
covariance behavior expressed by C. It is thus not surprising that a
full expression of covariance should include statements concerning the
measurement process, and O~II is such a statement. Another way to motivate
this assumptions is to argue that any measurement process, occurring as
it does in an interaction of a system with a macroscopic apparatus, is
already likely to involve space-like separated events. Thus any
physically reasonable system obeying C already must presuppose something
like O~II.

From \(s_2 = g\cdot s\) and Postulates \ref{pos:c} and \ref{pos:oi}, 
one has the following equality of
objective mixtures: 
\[
\sum_{i,j}\omega_{i,j}^{I \wedge J}(p)\pi^{I
\wedge J}_{i,j} p = \sum_j \sum_i \omega_j^J(\pi^{I}_i p) \omega_i^I(p)
\pi^{J}_j\pi^{I}_i p,
\] 
and thus, the final pure components and the
corresponding fractions must coincide. Assuming that the set of pure
states is sufficiently large to distinguish the functions \(\omega\) and
\(\pi\) one comes to the following consequence:

\begin{theorem}
Given  that \(I \x J\) where \(I
= (a_1,\dots,a_n)\) and \(J=(b_1,\dots,b_m)\)  then:
\begin{eqnarray}\label{eq:one}
\forall p \in \cP,\quad \omega_{i,j}^{I \wedge J}(p) &=&
\omega_j^J(\pi^I_i p)\omega_i^I(p) \\ \label{eq:two}
\pi^{I \wedge J}_{i,j} &=& \pi^J_j \pi^I_i.
\end{eqnarray}
\end{theorem}

These then are the constraints that any measurement theory of the type
described here must satisfy if it is to be able to describe
Lorentz-covariant physics.

One should note that in the two equations of the theorem, the left-hand
side is  the same if \(I\) and \(J\) are interchanged. This leads
immediately therefore to the following relations:
\begin{eqnarray} \label{eq:three}
\forall p \in \cP,\quad \omega_j^J(\pi^I_i p)\omega_i^I(p)
&= &
\omega_i^I(\pi^J_j p)\omega_j^J(p),\\ \label{eq:four}
 \pi^J_j \pi^I_i &=& \pi^I_i \pi^J_j.
\end{eqnarray}
These relations of course could have been independently derived from an
argument similar to ours if we considered two frames in which the
temporal order of experiments \(I\) and \(J\) are opposite, without
considering a frame in which they are simultaneous.

Now we come to the covering law. What is remarkable is that in various
axiomatic schemes that have been proposed to replace the covering
law with more ``physical" assumptions, these coincide with or are
readily deduced from one or both of equations (\ref{eq:one}--\ref{eq:two}) 
or even the weaker
results (\ref{eq:three}--\ref{eq:four}), 
with the difference that they are postulated for \(I
\leftrightarrow J\) and not only for \(I \x J\). We have been somewhat more
general in this exposition than what is normally assumed, as usually
the \(\omega_i^I(p)\) and \(\pi^I_i p\) are postulated not to depend
on \(I\) but only on the proposition \(a_i\) in question. This is the
absence of certain type of contextuality.

Bearing this in mind we thus see that the ``Compatibility Postulate" in
Guz \cite{guz1} is a direct consequence of (\ref{eq:two}) stated for
commuting pairs; axiom F4 in Guz \cite{guz2} follows from (\ref{eq:one}), 
again stated for commuting pairs, as can be found in
Svetlichny \cite{svet3}. Pool's \cite{pool1,pool2} derivation of the 
semimodularity of
quantum logic, necessary for a hilbert-space interpretation, follows a
different chain of reasoning but is clearly related to our results. The
axioms of the first paper  lead to the equivalence of
commutativity of propositions with a form of equation (\ref{eq:four}) 
while in the
second paper  semimodularity is derived from a further
assumption which in our scheme is M4. Nicolas Gisin (private
communications) has likewise derived the covering law from a form of
equation (\ref{eq:one}) stated for propensities and has 
independently postulated a
possible relation between the covering law and space-time structure.

Let us now consider the constraints that express the fact that local
observables must not distinguish between two objective mixtures not
related by an element of the group action but that are legitimately
maintained by two observers who disagree as to which experiments have
already been carried out. Suppose that in some region \(\cO\) an
experiment is performed which the first observer, space-like to \(\cO\),
says has already happened and consisted of an observation by an
instrument \(I\) upon a pure state \(p\). He would say the new state is
\(\sum_i\omega^I_i(p)\pi^I_ip\). Another observer flying past him would
claim that since the experiment has not been performed in his frame, the
state is \(g\cdot p\) where \(g\) is the element of the Lorentz group
that relates the two observers. Suppose now that a local experiment is
performed in a region \(\cO'\) space-like to \(\cO\). The first observer
describes this by an instrument \(J\) and the frequency of the \(j\)-th
result is thus given by \(\sum_i\omega^I_i(p)\omega^J_j(\pi^I_ip)\).
This by (3) is \(\sum_i\omega^I_i(\pi^J_jp)\omega^J_j(p) =
\omega^J_j(p)\). The second observer will assign frequency
\(\omega^{g\cdot J}_j(g\cdot p)\) to the same result but this by C is
also \(\omega^J_j(p)\) and so the two observers will agree as to the
frequencies of outcomes of local experiments and the two descriptions
are consistent just as in the quantum mechanical case. Finally we
mention that (\ref{eq:one}) also implies that no superluminal signal can
be sent through long-range correlation by mere change of a local
measuring device. The frequency assigned to the \(i\)-th result of
instrument \(I\), under lack of knowledge of the outcome of instrument
\(J\), is by the right-hand side of (\ref{eq:one}) just
\(\sum_j\omega_j^J(\pi^I_i p)\omega_i^I(p) = \omega^I_i(p)\) which is
independent of instrument \(J\) just as in the quantum-mechanical case
and so no signal of this type is possible.

\section{Bridging the Light Cone}

One sees that in some of the existing schemes the crucial axiom that
leads to the covering law can be substituted by Lorentz covariance and
an appeal to a principle generalizing necessary conditions on space-like separated
propositions to merely commuting ones. 
Let us first  examine the purely {\em logical\/}
 nature of such a principle which
can be stated as follows:
\begin{quote}(E -- Equivalence of commutativities) 
{\em Let \(Q(a,b)\) be two-place predicate concerning \(\cL\) then, \[(a \x b
 \Rightarrow Q(a,b)) \Rightarrow (a \leftrightarrow b \Rightarrow
 Q(a,b))\]}
\end{quote}

Assuming the metatheoretic principle E, the covering law can be
deduced from Lorentz covariance and axioms of generally less
controversial nature. The covering law would thus be compelling if one
can turn E compelling.
One has to be a bit careful though  about this principle, one should 
probably not
maintain it for all possible \(Q\). If for \(Q(a,b)\) 
we take \(a \x b\)
then we deduce \((a \leftrightarrow b) \Leftrightarrow (a \x b)\) which is
likely too strong. It may be that all commutativity in physics can be
reduced to space-like commutativity, but one should see this in detail
and not through a metatheoretic principle. 
 Is there reason to believe E? One way that E could be
true is if there is a symmetry group on \(\cL\) by which if \(a\) and \(b\)
are local and \(a
\leftrightarrow b\) then there is an element \(\phi\) of this group such
that \(\phi(a) \x \phi(b)\). This would establish E for group invariant
predicates and a join-dense set of propositions, which would probably be 
sufficient. In ordinary quantum
mechanics the full unitary group of hilbert space is apparently such a
symmetry group. There seems to be no compelling reason however to
believe in such a group. Given a complete lack of any
mathematical development of lattice-theoretic approaches ``localized"
into space-time regions and with Poincar\'e group action, ``local
relativistic quantum logic" in other words, it's even hard to say how
restrictive E is. From the physical side one lacks any real
understanding of commutativity of propositions except for space-like
separated ones where causality arguments seem compelling. Thus the spin
and orbital angular momenta of a particle commute, but we see this in a
purely formal way: the corresponding operators act on different tensor
factors. We cannot say we have a true physical understanding of this.
Without this understanding, the covering law is still not compelled.

To better understand what can be physically 
involved in assumptions such as E, it is necessary to examine more
closely the meaning of commutativity, which formalizes the notion of
compatibility. 
For the rest of this paragraph we assume a more primitive notion of an
instrument than that used in the previous section. 
An instrument \(I\) corresponds to some procedure or physical construct
which leads to a finite set of outcomes \(x_1,\dots,x_n\) but we do
not assume that these  are necessarily represented by some algebraic
elements. We can still talk about states \(p\), frequencies
\(\omega_i^I(p)\) and projections \(\pi^I_i p\) as before.
Operationally, we say that an instruments \(I\)  
with outcome set \(x_1,\dots,x_n\) and 
another one \(J\) with outcome set \(y_1,\dots,y_m\)
are compatible if there is a compound instrument \(K\) with outcome
set \((x_i,y_j),\, i=1,\dots,n,\, j=1,\dots,m\) whose outcome
frequencies in any state has as marginals the outcome frequencies of
\(I\) and \(J\) in the same state. The actual physical construction of
\(K\) may have in general no clear relation to the physical
constructions
of \(I\) and \(J\), but in certain cases it is customary to consider
\(K\) as just \(I\) and \(J\) physically coexisting and consider the
outcomes of \(K\) as coincidences of outcomes of \(I\) and \(J\). This
is the case when \(I\x J\) and also when \(I\) and \(J\) belong
to limited space-time regions in which all points of one are future
time-like to all points of the other.  In ordinary quantum mechanics,
the compound observation of compatible observations that succeed
temporally are taken to be just these successive individual
measurements. Two instruments that satisfy (\ref{eq:one}) for both
orders of \(I\) and \(J\) on the right hand side are compatible for then
the compound instrument  \(K\) is  \(I\wedge J\). 
Indeed one has for the two marginals:
\(\sum_j \omega_{i,j}^{I \wedge J}(p) = 
\sum_j\omega_j^J(\pi^I_i p)\omega_i^I(p) = \omega_i^I(p)\) and 
\(\sum_i \omega_{i,j}^{I \wedge J}(p) = 
\sum_i\omega_i^I(\pi^J_j p)\omega_j^J(p) = \omega_j^J(p)\).
For space-like separated instruments one can argue for both orders on
the right-hand side of (\ref{eq:one}) 
from covariance arguments as there are frames in
which the measurements occur in either temporal order. In the time-like
case one cannot use this argument. Let us consider the case in which
\(J\) belongs to a space-time region all points of which are
future time-like in relation to each point of the region to which \(I\)
belongs. One can then consider the instrument \(J\circ I\) 
consisting of the
successive measurements by \(I\) and then by \(J\). By our postulates
one has immediately \(\omega_{i,j}^{J\circ I}(p) = 
\omega_j^J(\pi^I_i p)\omega_i^I(p)\) and so (\ref{eq:one}) holds for one
order by definition even if \(J\) and \(I\) are not compatible. Now if
\(I\) and \(J\) {\em are\/} compatible, then conventional wisdom deems
\(J\circ I\) to be the compound instrument \(K\), and so 
half of (\ref{eq:one}) is trivially true. The difficulty then is in
arguing for the other half, that is, \(\omega_{i,j}^{J\circ I}(p) = 
\omega_i^I(\pi^J_j p)\omega_j^J(p)\). The operational verification of
this cannot now be simply done by successive measurements as the right
hand side now has an \(I\) measurement on a state conditioned by a
future time-like \(J\) measurement.  
The frequencies \(\omega_j^J(p)\) can be
established just from \(J\) measurements, but then to verify the 
frequencies \(\omega_i^I(\pi^J_j p)\) one needs to be able to create
states \(\pi^J_j p\) prior to the execution of \(I\), that is one needs
an instrument \(L\) with \(m\) outcomes, 
belongs to a region in the temporal past of
\(I\), whether time-like or space-like, and a state \(p'\) such that 
\(\pi^L_jp'=\pi^J_j p\). One can then verify (\ref{eq:one}). Thus 
the operational verification that this
equation holds for generally commutative instruments involves an
assumption concerning the producibility of given states by actions in
distinct and possibly widely separated regions
and so transcends purely covariant considerations. It seems that to
be able to perform the generalization mentioned in the beginning of this
section one needs a {\em physical\/} assumption going beyond covariance.

Standard relativistic quantum theory provides a clue by the presence of
long-range correlated states, that is, states of the EPR type. As a
simplified version of this situation suppose you want to study
right-hand circularly polarized photons. One way is to simply put an
appropriate filter in front of a light source and those photons that get
through are of the right kind and so can be observed at will. Another
equivalent way is to set up an EPR-type arrangement that creates singlet
two-photon states with the individual photons flying off in opposite
directions. Put now the same filter on the {\em distant\/} arm of the
EPR apparatus and {\em nothing\/} on the near arm. Observe at will. Half
of the photons observed are right-hand circularly polarized and half are
in the orthogonal left-hand circularly polarized state, and as the
measurements are done, there is no way of knowing which is which. If all
one wants however is analysis of experimental outcomes, this is no
problem, just wait enough time that the results (passage through the
filter or not) at the distant arm of each photon pair are available
(typical correlation experiment situation) and simply throw out all the
experimental data for the instances where the distant photon did not
pass through the filter. This provides you with data now of just the
right-hand circularly polarized photons at the near arm. The fact that
these two experimental procedures are equivalent is a feature of
ordinary quantum mechanics and depends on the existence of a particular
entangled state, the two-photon singlet. We now show that this
simplified situation is quite general. Assume we are working with a
local relativistic quantum field theory of the Haag type in a hilbert
space \(\cH\). Let \(I=(P_1,\dots,P_n)\) and \(J=(Q_1,\dots,Q_m)\) be
two compatible instruments belonging to two limited space-time regions,
\(\cO_I\) and \(\cO_J\) respectively, which need not be space-like
separated. Let \(\Psi\) be a pure state. Consider now an element \(g\)
of the Poincar\'e group such that \(\cO_J'=g(\cO_J)\) is space-like
separated from both the original regions. Let \(U(g)\) be the unitary
symmetry operator associated to \(g\), and let
\(J'=(Q_1',\dots,Q_m')\), where \(Q_j'=U(g)Q_jU(g)^*\), be the
transformed instrument. One can choose \(\cO_J'\) in such a 
manner \cite{haag, schroer} that \(\cH\) decomposes into a tensor product
\(\cH=\cH_1\otimes\cH_2\) with \(P_i = \tilde P_i\otimes I\), \(Q_j =
\tilde Q_j\otimes I\), and \(Q_j'=I\otimes \tilde Q_j'\). The projectors
of all three instruments commute among themselves. Let \(\Lambda\) be
the set of triples \(ijk\) such that \(P_iQ_jQ_k'\cH\neq \{0\}\), and
for \(ijk\in \Lambda\) let \(e_{ijk\alpha},\,\alpha\in A(ijk)\) be an
orthonormal  basis for \(P_iQ_jQ_k'\cH\). One has \(\Psi =
\sum_{ijk\in \Lambda}\sum_{\alpha\in A(ijk)}
\psi_{ijk\alpha}e_{ijk\alpha}\). Now for fixed \(ij\) one can find
an index \(p\) such that \(ipj\in \Lambda\), for otherwise one would have
\(\sum_pP_iQ_pQ_j'=P_iQ_j'=0\) which is impossible given the tensor
product decomposition. Choose \(\beta\in A(ipj)\), set 
\[\psi_{ipj\beta}'=\sqrt{\sum_{\{k\,|\,ijk\in \Lambda\}}
\sum_{\alpha\in A(ijk)}|\psi_{ijk\alpha}|^2}\]
and set all other components \(\psi_{ikj\alpha}'=0\) for \((k,\alpha)
\neq (p,\beta)\). Obviously 
\(\Psi' =
\sum_{ijk\in \Lambda}\sum_{\alpha\in A(ijk)}
\psi_{ijk\alpha}'e_{ijk\alpha}\) is another normalized state vector. One
easily verifies \(||P_iQ_k'\Psi'||^2=||P_iQ_k\Psi||^2\).
 In
standard quantum mechanics one has \(\omega_i^I(\Psi)=||P_i\Psi||^2\)
and \(\pi^I_i\Psi=P_i\Psi/||P_i\Psi||\), whenever
\(\omega_i^I(\Psi)\neq0\). 
We have from our construction that \(\omega_{i,j}^{I
\wedge J'}(\Psi')=\omega_{i,j}^{I \wedge J}(\Psi)\),
\(\omega_j^{J'}(\pi^I_i \Psi')=\omega_j^J(\pi^I_i \Psi)\) and
\(\omega_i^I(\Psi') =\omega_i^I(\Psi)\). So the validity of
(\ref{eq:one}) for the pair of instruments \((I,J)\) can be deduced
from the validity for the pair \((I,J')\) of {\em space-like
separated\/} instruments {\em due to the existence of 
the state \(\Psi'\)\/}.
Note that no distinction is made between the two instruments so that the
right-hand side of (\ref{eq:one}) holds for any order. 
The validity of (\ref{eq:one}) for space-like separated instruments
follows basically from Lorentz covariance and other assumptions of this
paper and so is not peculiar to standard quantum theory. The existence
of \(\Psi'\) has to be viewed however as an additional physical
assumption, that happens to be true in standard quantum theory and is in
some sense also characteristic of it. 
This state is in general one that has long-range correlations of
the EPR-type. We can now envisage a more physically plausible version of 
principle E:
\begin{postulate}[EP - Equivalence of conditioning] For any pair of
instruments \((I,J)\)  with \(I\leftrightarrow J\) and pure state \(p\) 
there is an instrument \(J'\) such that \(J' \x I\), and a pure state
\(p'\) such that the joint experiment \((I,J)\) on the state \(p\) is
equivalent to the joint experiment \((I,J')\) on \(p'\). That is, 
\(\omega_{i,j}^{I\wedge J'}(p')=\omega_{i,j}^{I \wedge J}(p)\), 
\(\omega_j^{J'}(\pi^I_i p')=\omega_j^J(\pi^I_i p)\) and 
\(\omega_i^I(p') =\omega_i^I(p)\)
\end{postulate}

We call this equivalence of conditioning since it allows one to prepare
states conditioned to outcomes at space-like separations as was the case
for the simplified photon polarization experiment. With this additional postulate, as
was pointed out before, one can now deduce the covering law
in some of the existing
axiomatization schemes.

\section{Conclusions}

One of the most intriguing features of quantum mechanics is its
universality. All phenomena, to the extent that their quantum behavior
can be exhibited experimentally, are subject to the same general
formalism. The postulated connection of quantum mechanics to space-time
structure makes this understandable. The measurements to which the above
discussion refer could be {\em any} measurements. To the extend that any
measurement takes place in space-time, it must exhibit universal quantum
behavior. The universality of quantum mechanics is a reflection of the
universality of space-time as the arena for our experiments.

To be able to deduce such universality one however has to be able to
make some assumption such as EP compelling. Now why should EP be
compelling? It is unlikely that one can find an argument on purely
formal grounds or by appeal to ``reasonableness" of any kind just as
such appeals are ultimately unconvincing in all the axiomatic approaches
to quantum mechanics. Space-like and time-like situations are in
logically distinct domains and any relation between them must come from
some realm in which this distinction is weakened. The only existing
considerations of this sort come from what is loosely known as ``quantum
gravity". In fact, at this point one perceives a fundamental difficulty
with the whole argument of this paper. One has started with a definitely
classical view of space-time and traced out a route which leads to a
universal mechanics. But now space-time phenomena themselves must, by
universality, be subject to the same mechanics, which distorts the
original starting point. Space-time itself must be quantum mechanical.
There is a self-consistency question. The constraints that space-time
structure places on mechanics must govern the structure of space-time
itself. Unfortunately ``quantum-space-time quantum logic" is just a
glimmer of an idea at this moment, more remote than the ``relativistic
quantum logic" that we've embarked upon. In any case, in all present-day
approaches to quantum gravity, the rigid structure of the light cone
disappears and the usual notions of  space-like and time-like are
emergent and not fundamental. In such a context it is perfectly
understandable that relations between space-like and time-like
situations arise out of a more basic  theory in which such a
distinction is not fundamental. 
 
On a more technical side, we note that the above argument utilizes more
the lorentzian causal structure of space-time and the relativity of
simultaneity than the exact details of Lorentz covariance. Thus one can
expect that a similar result can be obtained for curved space-time as
well. Strict Lorentz covariance should therefore be viewed as a
simplifying assumption for these preliminary studies. One also has the
awkwardness of deriving a global feature (covering law) through local
considerations. One knows from algebraic quantum field theory that local
von Neumann algebras of observables have a unique form (Haag \cite{haag}), 
they
are all type \(\hbox{III}_1\) hyperfinite factors for a causal diamond
\(\cO\) (intersection of a forward and a backward light cone). It would be
more in keeping with the local approach to try to deduce that \(\cL(\cO)\)
is isomorphic to the projection lattice of a type \(\hbox{III}_1\)
hyperfinite factor and then only secondarily argue for the covering law
through global considerations. This of course requires a
lattice-theoretic characterization of such factors, which to our
knowledge is not available.

\section{Acknowledgment}

The author thanks professor Nicolas Gisin for helpful correspondence.
Special thanks go to the Mathematics Department of Rutgers University
for its hospitality during the author's stay there where part of this
work was done. This research was financially supported by the Secretaria
de Ci\^encia e Tecnologia (SCT) and the Conselho Nacional de
Desenvolvimento Cient\'\i fico e Tecnol\'ogico (CNPq), both agencies of
the Brazilian government.


\end{document}